\documentclass[10pt]{article}
\usepackage[]{amsmath}
\usepackage[]{amssymb}
\usepackage[]{amsthm}
\usepackage{bm}
\usepackage[]{geometry}
\usepackage{graphicx}
\usepackage[]{natbib}
\usepackage{color}
\usepackage{array}
\usepackage{url}
\usepackage{paralist}
\usepackage{rotating}
\usepackage{hhline}
\usepackage{authblk}
\usepackage{setspace}

\DeclareMathOperator*{\argmax}{arg\,max}

\DeclareMathOperator*{\cov}{cov}

\begin{document}
	
\title{Bias in multivariable Mendelian randomization studies due to measurement error on exposures}
\author[1]{Jiazheng Zhu}
\author[1,2]{Stephen Burgess}
\author[1]{Andrew J. Grant\thanks{Corresponding author. Email address: andrew.grant@mrc-bsu.cam.ac.uk}}
\affil[1]{\normalsize MRC Biostatistics Unit, University of Cambridge, Cambridge, UK}
\affil[2]{\normalsize Cardiovascular Epidemiology Unit, University of Cambridge, Cambridge, UK}
\date{}

\maketitle

\begin{abstract}
	Multivariable Mendelian randomization estimates the causal effect of multiple exposures on an outcome, typically using summary statistics of genetic variant associations. However, exposures of interest in Mendelian randomization applications will often be measured with error. The summary statistics will therefore not be of the genetic associations with the exposure, but with the exposure measured with error. Classical measurement error will not bias genetic association estimates but will increase their standard errors. With a single exposure, this will result in bias toward the null in a two sample framework. However, this will not necessarily be the case with multiple correlated exposures. In this paper, we examine how the direction and size of bias, as well as coverage, power and type I error rates in multivariable Mendelian randomization studies are affected by measurement error on exposures. We show how measurement error can be accounted for in a maximum likelihood framework. We consider two applied examples. In the first, we show that measurement error leads to the effect of body mass index on coronary heart disease risk to be overestimated, and that of waist-to-hip ratio to be underestimated. In the second, we show that the proportion of the effect of education on coronary heart disease risk which is mediated by body mass index, smoking and blood pressure may be underestimated if measurement error is not taken into account.
\end{abstract}


\section{Introduction}
Mendelian randomization is a method for estimating the effect of an exposure on an outcome using genetic variants as instrumental variables \citep{Lawlor2008}. By using genetic variation as a proxy, or instrument, for changes in the exposure, its effects on the outcome can be estimated free of environmental factors which typically confound the exposure-outcome relationship. For a genetic variant to be a valid instrumental variable, it must satisfy the following three assumptions: it must be associated with the risk factor; it must be independent of any confounders of the risk factor-outcome relationship; and it must be independent of the outcome conditional on the risk factor and confounders \citep{Greenland2000}. Using genetic instruments which satisfy these assumptions, Mendelian randomization can provide evidence for a causal effect of the exposure on the outcome.

The Mendelian randomization paradigm can be extended to include multiple exposures, known as multivariable Mendelian randomization \citep{BThompson2015mv}. Violations of the third instrumental variables assumption due to association pathways between the genetic variants and the outcome via measured traits can be accounted for by including these traits in a single model. Furthermore, the multivariable framework can disentangle the causal effect of an exposure into its direct effect on the outcome and its indirect effects via other measured traits \citep{Burgess2017dissecting}. For a genetic variant to be a valid instrumental variable in a multivariable analysis it must satisfy the following three assumptions: it must be associated with at least one risk factor; it must be independent of confounders of all exposure-outcome relationships; and it must be independent of the outcome conditional on all exposures and confounders \citep{Sanderson2019}.

It is common in Mendelian randomization studies that individual level data is not available, only summary statistics from genome-wide association studies \citep{BButterworthThompson2013}. Often, the genetic variant association estimates with the exposures are taken in separate samples to those with the outcome, a framework known as two-sample Mendelian randomization \citep{Hartwig2018}. The most common technique for estimating the causal effects of the exposures on the outcome using summary level data is to perform weighted regression of the genetic variant-outcome association estimates on the genetic variant-exposure association estimates. The inverse-variance weighted method (IVW) uses as weights the inverse of the standard errors of the genetic variant-outcome association estimates \citep{BButterworthThompson2013, BDudbridgeThompson2015mv}.

Exposures of interest in Mendelian randomization studies may not be easy to measure precisely, and thus may be subject to measurement error. The inputs to the IVW model will therefore not be estimates of the associations of the genetic variants with the exposures, but rather with the exposures measured with some error. \citet{PierceVanderWeele2012} categorise measurement error into systematic and classical: systematic measurement error biases association estimates and is a result of, for example, miscalibration of measurements; classical measurement error is where errors are randomly distributed around the true value and independent of the trait being measured. The latter type will not bias genetic association estimates, but will increase their standard errors. The IVW approach ignores any uncertainty in the genetic variant-exposure association estimates, thus estimates computed in this way will not account for classical measurement error. In the single exposure, or univariable, case, it has been shown that uncertainty in the genetic variant-exposure estimates will bias IVW estimates in the two-sample framework toward the null \citep{PierceBurgess2013}. The IVW approach is therefore considered conservative, in that measurement error will reduce the power of the method to detect a true effect, but still provides a valid test of the causal null hypothesis. However, this is not necessarily true in the multiple risk factor case \citep{Sanderson2019}.

In this paper we examine the effect of classical measurement error, which from hereon in we shall refer to simply as measurement error, on multivariable Mendelian randomization estimates, which has not previously been explored in detail. We focus on two-sample Mendelian randomization using summary statistics of genetic associations with multiple exposures. We also show how measurement error can affect estimation in mediation analyses. We discuss strategies for dealing with measurement error and show that accounting for uncertainty in the genetic variant-exposure association estimates in a maximum likelihood framework produces unbiased estimates with nominal coverage and type I error rates. Finally, we consider two applied examples. The first examines the effects of body mass index and waist-to-hip ratio on the risk of coronary heart disease. The second considers the proportion of the effect of educational attainment on coronary heart disease which is mediated by body mass index, smoking and blood pressure.

\section{Modeling assumptions} \label{se:model}
\begin{figure}
	\centering
	\includegraphics{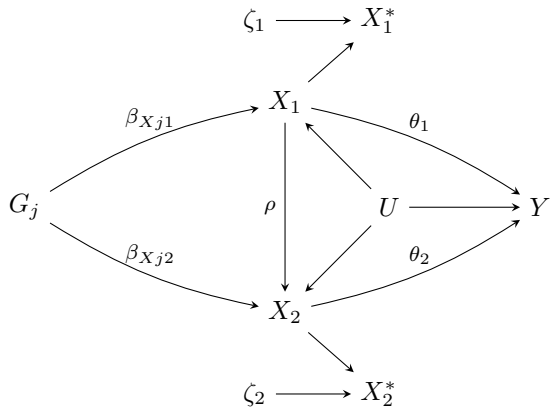}
	\caption{Directed acyclic graph showing the relationship between the $j$th genetic variant ($G_{j}$), two exposures ($X_{1}, X_{2}$), an outcome ($Y$) and unmeasured confounders ($U$). The observed exposures ($X_{1}^{*}, X_{2}^{*}$) are subject to measurement error ($\zeta_{1}$, $\zeta_{2}$).}
	\label{fg:dag}
\end{figure}

We consider the case where we have $K$ exposures, denoted by $X_{1}, \ldots, X_{K}$, and we are interested in their causal effects, denoted by $\theta = \left( \theta_{1}, \ldots, \theta_{K} \right)$, on an outcome $Y$. We assume that the relationship between the exposures and the outcome are linear and homogenous (that is, no effect modification). The association between the exposures and the outcome are potentially confounded by unmeasured variables which we represent as the single variable $U$. We have $J$ genetic variants, $G_{1}, \ldots, G_{J}$, which we assume satisfy the three instrumental variables assumptions. We further assume that the genetic variants are independent of each other and that $J > K$. The exposures may also be causally related, such as in the case where one is a mediator of the relationship between another exposure and the outcome. This scenario is illustrated for the two exposure case in Figure \ref{fg:dag}.

We suppose that we do not have access to the individual level data, only summary statistics of the marginal associations between each genetic variant and the exposures and outcome. Let $\hat{\beta}_{Yj}$ be the estimated association between the $j$th genetic variant and the outcome and let $\sigma_{Yj}$ be its standard error. Although $\sigma_{Yj}$ is estimated from data, we assume it is known. Further, let $\beta_{Xj}$ be the vector of associations between the $j$th genetic variant and the $K$ exposures. Note that under the summary level Mendelian randomization framework, the traits may be either continuous or binary. If they are continuous, the associations are typically estimated using simple linear regression. If they are binary, they are typically estimated using logistic regression. Finally, we assume that the $J \times K$ matrix with $j$th row $\beta_{Xj}'$, denoted $\beta_{X}$, is of full column rank. Under the given assumptions, $\theta_{k}$ represents the direct causal effect of $X_{k}$ on $Y$ \citep{Sanderson2019}.

\section{Estimating the causal effects} \label{se:estimation}

\subsection{The inverse-variance weighted method} \label{se:ivw}
If the $j$th genetic variant is a valid instrumental variable then $\beta_{Yj} = \beta_{Xj}' \theta$. Given the large sample sizes in which the genetic associations are typically estimated in, it is reasonable to approximate the distribution of $\hat{\beta}_{Yj}$ as normal with mean $\beta_{Xj}' \theta$ and variance $\sigma_{Yj}^{2}$. A standard assumption in Mendelian randomization is that the estimates of the genetic associations with the exposures have zero standard error, which is equivalent to these associations being estimated with an infinite sample size \citep{Bowden2017}. We shall therefore denote the observed genetic associations with the $j$th exposure by $\beta_{Xj}$, reflecting the fact that this is assumed to be the true association value. This motivates the IVW method for estimating $\theta$, which fits the weighted linear regression model
\begin{equation}
	\hat{\beta}_{Yj} = \beta_{Xj}' \theta + \varepsilon_{Yj} , \label{eq:ivwmodel}
\end{equation}
$j = 1, \ldots, J$, where $\varepsilon_{Yj}$ is an error term, independent of $\beta_{Xj}$, which is normally distributed with mean $0$ and variance $\sigma_{Yj}^{2}$. The MR-IVW estimator is thus
\begin{equation}
	\hat{\theta}_{\textrm{IVW}} = \left( \sum_{j=1}^{J} \sigma_{Yj}^{-2} \beta_{Xj} \beta_{Xj}' \right)^{-1} \sum_{j=1}^{J} \sigma_{Yj}^{-2} \hat{\beta}_{Yj} \beta_{Xj} . \label{eq:est_ivw}
\end{equation}
Substituting (\ref{eq:ivwmodel}) into (\ref{eq:est_ivw}), we have
\begin{equation}
	\hat{\theta}_{\textrm{IVW}} - \theta = \left( \sum_{j=1}^{J} \sigma_{Yj}^{-2} \beta_{Xj} \beta_{Xj}' \right)^{-1} \sum_{j=1}^{J} \sigma_{Yj}^{-2} \varepsilon_{Yj} \beta_{Xj} . \label{eq:ivwbias}
\end{equation}
Due to the independence of $\beta_{Xj}$ and $\varepsilon_{Yj}$, $E \left( \hat{\theta}_{\textrm{IVW}} - \theta \right) = 0$ and so $\hat{\theta}_{\textrm{IVW}}$ is an unbiased estimator of $\theta$.

\subsection{Exposures measured with error} \label{se:measurement_error}
Suppose we do not have the genetic variant associations with $X_{k}$, but rather with $X^{*}_{k} = X_{k} + \zeta_{k}$, where $\zeta_{k}$ is a random variable with mean zero, independent of $X_{k}$, $G_{1}, \ldots, G_{J}$ and $\zeta_{l}$, $l \neq k$. The association estimates for the $j$th genetic variant with the measured exposures are thus
\begin{equation}
	\hat{\beta}_{Xj}^{*} = \beta_{Xj} + \hat{\beta}_{\zeta j}, \label{eq:betaXstar}
\end{equation}
where $\hat{\beta}_{\zeta j}$ is a vector of independent random variables where the $k$th element has mean zero and variance which is proportional to the variance of $\zeta_{k}$. Denoting by $\hat{\theta}_{\textrm{IVW}}^{*}$ the IVW estimator using the genetic association estimates with the measured exposures, we have
\begin{equation}
	\hat{\theta}_{\textrm{IVW}}^{*} - \theta = \left( \sum_{j=1}^{J} \sigma_{Yj}^{-2} \hat{\beta}^{*}_{Xj} \hat{\beta}^{*'}_{Xj} \right)^{-1} \sum_{j=1}^{J} \sigma_{Yj}^{-2} \hat{\beta}^{*}_{Xj} \left(  \varepsilon_{Yj} - \hat{\beta}_{\zeta j}' \theta \right) . \label{eq:ivw_me_bias}
\end{equation}
If the exposures are measured without error, that is, if $\hat{\beta}_{X}^{*} = \beta_{X}$, then we recover (\ref{eq:ivwbias}) and the estimator is unbiased. However, if measurement error is non-zero, then the bias will be non-zero and can be in any direction with respect to the null.

From hereon in we focus on the $K=2$ case, although our results can be extended to higher dimensions. Assume, without loss of generality, that the columns of $\beta_{X}$ are centered to have mean zero. Define, for $k, l = 1, 2$,
\begin{equation}
J^{-1} \sum_{j=1}^{J} \sigma_{Yj}^{-2} \hat{\beta}_{\zeta j k} \hat{\beta}_{\zeta j l} \rightarrow_{p} \left\lbrace  \begin{array}{l l} v_{\zeta k}, & k=l \\ 0, & \textrm{otherwise} \end{array} \right. \label{eq:var_zeta}
\end{equation}
and
\begin{equation}
J^{-1} \sum_{j=1}^{J} \sigma_{Yj}^{-2} \hat{\beta}_{X j k}^{*} \hat{\beta}_{X j l}^{*} \rightarrow_{p} \left\lbrace  \begin{array}{l l} v_{X k}^{*}, & k=l \\ c_{X}^{*}, & \textrm{otherwise} \end{array} \right. , \label{eq:var_betaXstar}
\end{equation}
where $\rightarrow_{p}$ denotes convergence in probability as $J \rightarrow \infty$. Let
\[
	\lambda_{k} = \frac{v_{\zeta k}}{v_{X k}^{*}}, \quad \rho^{*} = \frac{c_{X}^{*}}{\sqrt{v_{X1}^{*} v_{X2}^{*}}} .
\]
The parameters $\lambda_{1}, \lambda_{2}$ quantify the level of measurement error, and will be zero if the exposures are measured precisely. The parameter $\rho^{*}$ quantifies the correlation between the genetic variant associations with the two measured traits. We have
\begin{align}
\hat{\theta}_{IVW, 1}^{*} - \theta_{1} &\rightarrow_{p} - \frac{\lambda_{1} \theta_{1} - \lambda_{2} \rho^{*} \sqrt{\frac{v_{X2}^{*}}{v_{X1}^{*}}} \theta_{2}}{1 - \rho^{*2}} \label{eq:ivw1_bias}\\
\hat{\theta}_{IVW, 2}^{*} - \theta_{2} &\rightarrow_{p} - \frac{\lambda_{2} \theta_{2} - \lambda_{1} \rho^{*} \sqrt{\frac{v_{X1}^{*}}{v_{X2}^{*}}}  \theta_{1}}{1 - \rho^{*2}}. \label{eq:ivw2_bias}
\end{align}
(see Section S1 in the Supporting Information). Note that this agrees with the established literature on measurement error in linear regression models (see, for example, \citealp{Maddala1992}).

It follows from (\ref{eq:ivw1_bias}) that if $K=1$, the bias due to measurement error will tend to be toward the null, which agrees with the established literature for IVW estimates in the single exposure case. In contrast, in the multiple exposure case, the bias due to measurement error can be in any direction, and depends on the level of measurement error as well as the correlation between the genetic association estimates. Only in the case where the correlation between the genetic association estimates is zero (that is, where $\rho^{*} = 0$) will the bias necessarily tend toward the null. Furthermore, measurement error on one exposure may bias the effect estimate for another exposure which is measured precisely. Suppose, for example, that $\theta_{1} \neq 0$, $\theta_{2} = 0$,  $\rho^{*} > 0$, and that $X_{1}$ is measured with error but $X_{2}$ is measured precisely. The estimate of $\theta_{2}$ will tend to be biased away from the true null value, and this bias may lead to incorrect conclusions that $X_{2}$ is a causal risk factor for $Y$ but $X_{1}$ is not.

\subsection{Accounting for measurement error on exposures} \label{se:methods}

\subsubsection{Simulation extrapolation}
The method of simulation extrapolation (SIMEX) has been proposed to account for measurement error in the univarable Mendelian randomization setting \citep{Bowden2016I2}. The approach works in two stages. In the first stage, sets of genetic variant-exposure associations are generated from the normal distribution with mean $\beta_{Xj}$ and variance $\left( 1 + \lambda \right) \sigma_{Xj}^{2}$, $j = 1, \ldots, J$, for successive values of $\lambda > 0$. Here, $\sigma_{Xj}$ is the standard error of the $j$th genetic variant-exposure association estimate from GWAS summary statistics. For each simulated set of $\beta_{Xj}$ values, $\theta$ is estimated using the observed $\hat{\beta}_{Yj}$ values. For each value of $\lambda$, the mean estimates of $\theta$ are taken. The second stage then extrapolates these mean values back to where $\lambda = -1$, which gives an estimate of $\theta$ when there is no measurement error. This extrapolation is done by fitting a model, such as a linear or quadratic model, to the mean values. Standard errors are computed using a jackknife approach \citep{CookStefanski1994, StefanskiCook1995}.

Although SIMEX has been shown to perform well for univariable Mendelian randomization, the approach becomes problematic in the multivariable case. Each exposure which is suspected to be subject to measurement error would need a separate $\lambda$. Even in a low-dimensional setting, it may be difficult to find an appropriate extrapolation model with multiple $\lambda$ parameters. Furthermore, we show in the following sections how a fast and easily implemented algorithm motivated by a maximum likelihood approach can effectively account for measurement error. We shall therefore not proceed with the SIMEX approach for the multivariable setting.

\subsubsection{A maximum likelihood approach} \label{se:mle}
From the model specified in Sections \ref{se:model}, \ref{se:ivw} and \ref{se:measurement_error}, for $j=1, \ldots, J$, $\hat{\beta}_{Yj}$ is normally distributed with mean $\beta_{Xj}' \theta$ and variance $\sigma_{Yj}^{2}$, and $\hat{\beta}_{Xj}^{*}$ is normally distributed with mean $\beta_{Xj}$ and covariance $\Sigma_{X j}^{*} = \cov \left( \hat{\beta}_{Xj}^{*} \right)$. If we assume the $\Sigma_{X j}^{*}$ matrices are known, the log-likelihood function, up to an additive constant, is thus
\begin{equation}
l \left( \theta, \beta_{X} \right) = - \frac{1}{2} \sum_{j=1}^{J} \left\lbrace  \left( \hat{\beta}_{Xj}^{*} - \beta_{Xj} \right)' \Sigma_{X j}^{*-1} \left( \hat{\beta}_{Xj}^{*} - \beta_{Xj} \right) + \sigma_{Yj}^{-2} \left( \hat{\beta}_{Yj} - \beta_{Xj}' \theta \right)^{2} \right\rbrace . \label{eq:lik}
\end{equation}
The parameter of interest is $\theta$ and the remaining unknown parameters, $\beta_{X1}, \ldots, \beta_{XJ}$, are considered nuisance parameters. If we fix $\theta$ and maximise $l \left( \theta, \beta_{X} \right)$ with respect to $\beta_{X}$, we obtain the profile likelihood
\begin{equation}
\tilde{l} \left( \theta \right) = l \left\lbrace \theta, \argmax_{\beta_{X}} l \left( \theta, \beta_{X} \mid \theta \right) \right\rbrace = - \frac{1}{2} \sum_{j=1}^{J} \frac{1}{\sigma_{Yj}^{2} + \theta' \Sigma_{X j}^{*} \theta} \left( \hat{\beta}_{Yj} - \hat{\beta}_{Xj}^{* \prime} \theta \right)^{2} . \label{eq:prolik}
\end{equation}

There is no closed form solution to (\ref{eq:prolik}), but it can be solved using any optimisation procedure. We propose an iterative approach. For fixed values of the $\beta_{Xj}$'s, (\ref{eq:lik}) can be easily maximised with respect to $\theta$. Similarly, for a given value of $\theta$, (\ref{eq:lik}) can be easily maximised with respect to the $\beta_{Xj}$'s. Our proposed procedure is thus as follows. Given estimates of the $\beta_{Xj}$'s, denoted $\tilde{\beta}_{Xj}$, we estimate $\theta$ from
\[
\hat{\theta} =  \left( \sum_{j=1}^{J} \sigma_{Yj}^{-2} \tilde{\beta}_{Xj} \tilde{\beta}_{Xj}' \right)^{-1} \sum_{j=1}^{J} \sigma_{Yj}^{-2} \hat{\beta}_{Yj} \tilde{\beta}_{Xj} .
\]
We then update the estimates of $\beta_{Xj}$ by
\[
\tilde{\beta}_{Xj} = \left( \sigma_{Yj}^{-2} \hat{\theta} \hat{\theta}' + \Sigma_{X j}^{*-1} \right)^{-1} \left( \sigma_{Yj}^{-2} \hat{\beta}_{Yj} \hat{\theta} + \Sigma_{X j}^{*-1} \hat{\beta}_{Xj}^{*} \right) ,
\]
$j = 1, \ldots, J$. We then re-estimate $\theta$, and the procedure continues until the difference in $l \left( \hat{\theta} \right)$ for successive iterations is below some predetermined threshold. For initial values of $\tilde{\beta}_{Xj}$, we randomly sample from the normal distribution with mean $\hat{\beta}_{Xj}^{*}$ and variance $\Sigma_{Xj}^{*}$. Confidence intervals for $\theta$ can be constructed using an asymptotic variance (see Section S2 in the Supporting Information).

In order to implement this approach in practice, the $\Sigma_{X j}^{*}$ matrices are required. The diagonal entries of this matrix can be taken from GWAS summary statistics. However, the off-diagonal entries will not always be known. In the case where the genetic associations with each exposure are uncorrelated, the off-diagonal entries of these matrices will be zero. When there is such correlation, the off-diagonal entries can be estimated if estimates of the correlation between the exposures are available \citep{Sanderson2021}. If these correlation estimates are not available, there are also methods for estimating the genetic association estimate covariances using summary statistics \citep{BulikSullivan2015, Ray2018}.

\subsection{Measurement error in mediation analysis}
One context in which multivariable Mendelian randomization can be used is to assess the degree to which the effect of an exposure on an outcome is mediated by other measured variables. If potential mediators are included in a multivariable Mendelian randomization analysis, then the estimand is the direct effect of the exposure on the outcome excluding any effects via pathways which include these mediators. We refer to this as the direct effect \citep{Burgess2017dissecting}. In contrast, the estimand in a univariable Mendelian randomization, using genetic variants which satisfy the univariable instrumental variables assumptions, is the total effect of the exposure on the outcome, which includes any mediated pathways. By comparing estimates of the direct and total effects, we can assess the relative contribution of causal pathways from the exposure to the outcome via other measured exposures in a mediation analysis.

For example, Figure \ref{fg:dag} illustrates the scenario where $X_{2}$ is a mediator of the effect of exposure $X_{1}$ on outcome $Y$. The total effect of $X_{1}$ on $Y$, estimated by univariable Mendelian randomization, is $\theta_{1} + \rho \theta_{2}$. The direct effect of $X_{1}$ on $Y$, estimated by multivariable Mendelian randomization, is $\theta_{1}$. The mediated effect of $X_{1}$ on $Y$ relative to the total effect is thus
\begin{align*}
	\frac{\rho \theta_{2}}{\theta_{1} + \rho \theta_{2}} &= \frac{\textrm{Total effect} - \textrm{Direct effect}}{\textrm{Total effect}}\\
		&= 1 - \frac{\textrm{Direct effect}}{\textrm{Total effect}} .
\end{align*}
This quantity can therefore be estimated by $1$ minus the ratio of the estimate of the effect of the exposure on the outcome from a multivariable analysis to that from a univariable analysis. If the direct and mediated effects are in the same direction, then this quantity may be thought of as the proportion of the effect of the exposure on the outcome which is mediated via other measured variables.

Bias caused by measurement error will thus bias an estimate of the relative effect of the exposure mediated by other variables. For example, measurement error on the exposure variable will tend to bias estimates of the total effect toward the null. As shown by (\ref{eq:ivw1_bias}), if the direct and mediated effects are in the same direction, then measurement error on both exposure and mediator will tend to bias estimates of the direct effect away from the null. In this scenario, the estimated proportion of the effect mediated by the secondary exposure will tend to be underestimated.

\section{Simulation study} \label{se:sims}

\subsection{Causal effect estimation}
\begin{figure}
	\centering
	\includegraphics{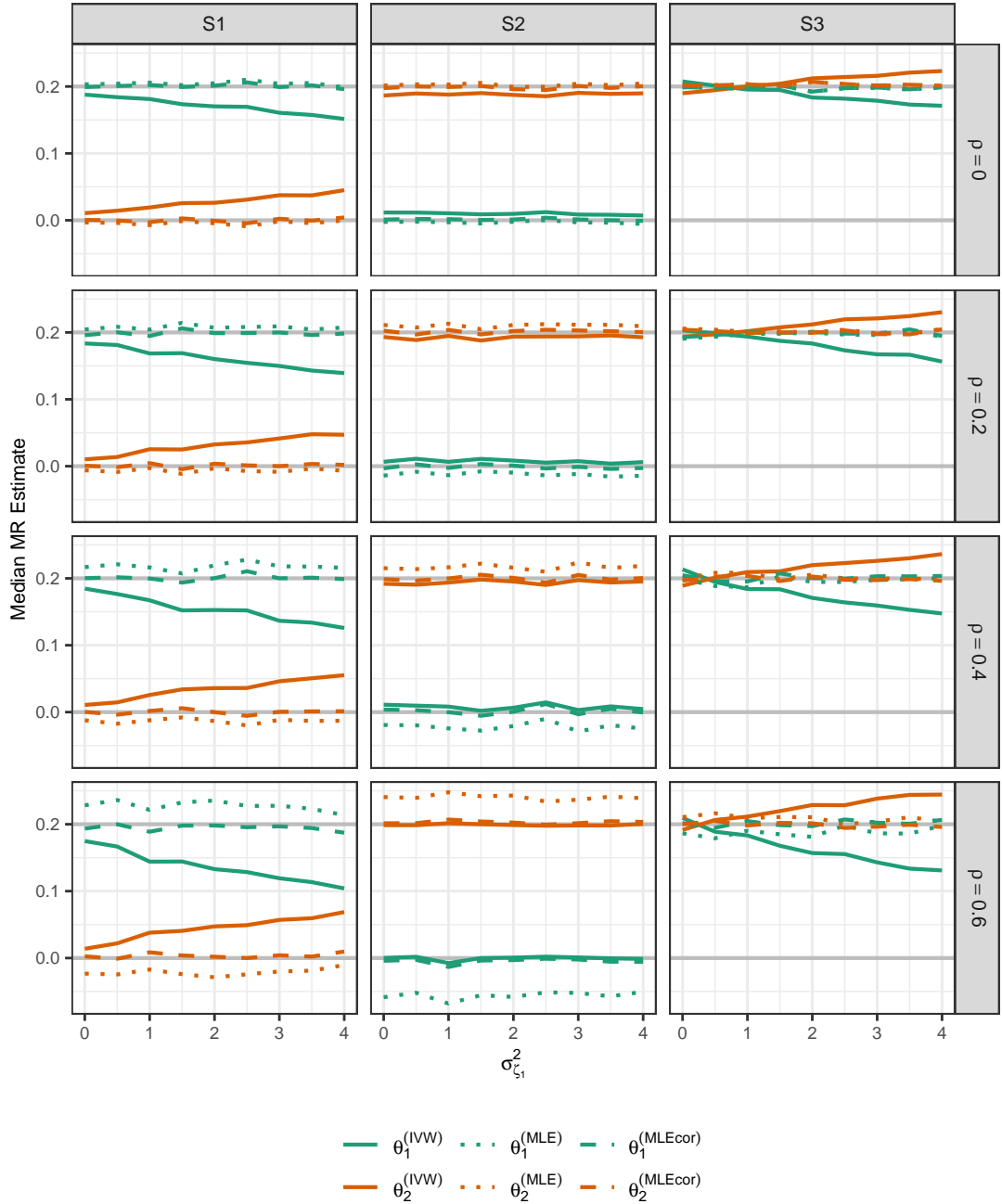}
	\caption{Median estimates of the causal effects from simulation Scenarios 1--3, for varying values for the causal effect of $X_{1}$ on $X_{2}$ ($\rho$) and the variance of the measurement error on $X_{1}$ ($\sigma_{\zeta_{1}}^{2}$), using IVW ($\theta_{1}^{\textrm{IVW}}$, $\theta_{2}^{\textrm{IVW}}$) and maximum likelihood estimation without ($\theta_{1}^{\textrm{MLE}}$, $\theta_{2}^{\textrm{MLE}}$) and with ($\theta_{1}^{\textrm{MLEcor}}$, $\theta_{2}^{\textrm{MLEcor}}$) sample trait correlation included. The thick grey lines indicate $0.2$ and $0$, which are the true causal effects in the various settings.}
	\label{fg:bias}
\end{figure}

In order to demonstrate the effect of measurement error on exposures on multivariable IVW estimates, we generated data for $n = 20\,000$ individuals and $J = 40$ genetic variants from the following model.
\begin{align*}
X_{1} &= \sum_{j=1}^{J} G_{j} \beta_{Xj1} + \gamma U + \sqrt{1 - \gamma^2} \varepsilon_{X1} \\
X_{2} &= \rho X_{1} + \sqrt{1 - \rho^2} \left( \sum_{j=1}^{J} G_{j} \beta_{Xj2} + \gamma U + \sqrt{1 - \gamma^2} \varepsilon_{X2} \right) \\
X_{1}^{*} &= X_{1} + \sigma_{\zeta_{1}} \zeta_{1} \\
X_{2}^{*} &= X_{2} + \sigma_{\zeta_{2}} \zeta_{2} \\
Y &= \theta_{1} X_{1} + \theta_{2} X_{2} +  U + \varepsilon_{Y} ,
\end{align*}
where, independently for $j=1, \ldots, J$, $k = 1, 2$,
\begin{align*}
\beta_{Xjk} &\sim \textrm{Uniform} \left( 0.08, 0.2 \right) , \\
G_{j} &\sim \textrm{Binomial} \left( 2, \textrm{maf}_{j} \right) , \\
\textrm{maf}_{j} &\sim \textrm{Uniform} \left( 0.01, 0.5 \right) , \\
U, \varepsilon_{Xk}, \zeta_{k}, \varepsilon_{Y} &\sim N \left( 0, 1 \right) .
\end{align*}
In this model, the correlation between genetic association estimates for the two exposures is affected by the $\rho$ and $\gamma$ parameters. We set $\gamma = 0.2$ in all simulations so that the contribution to the correlation between the traits from shared effects of the confounder $U$ is relatively small and the correlation between genetic association estimates is thus approximately equal to $\rho$. We varied $\rho$ between $0$ and $0.6$. The amount of measurement error on $X_{1}$ was increased by varying $\sigma_{\zeta_{1}}^{2}$ between $0$ and $4$. We considered the following three scenarios.
\begin{enumerate}
	\item[S1.] $X_{1}$ was causal ($\theta_{1} = 0.2$), $X_{2}$ was not causal ($\theta_{2} = 0$) and was measured without error ($\sigma_{\zeta_{2}}^{2} = 0$).
	\item[S2.] $X_{1}$ was not causal ($\theta_{1} = 0$), $X_{2}$ was causal ($\theta_{2} = 0.2$) and was measured without error ($\sigma_{\zeta_{2}}^{2} = 0$).
	\item[S3.] $X_{1}$ and $X_{2}$ were both causal ($\theta_{1} = \theta_{2} = 0.2$) and $X_{2}$ was measured with error ($\sigma_{\zeta_{2}}^{2} = 1$).
\end{enumerate}
Genetic association estimates were computed using simple linear regression for $X_{1}^{*}$, $X_{2}^{*}$ and $Y$ on each $G_{j}$. For each of $1\,000$ replications, two separate samples were generated, with the genetic associations with the exposures estimated in one, and with the outcome in the other. We applied three methods to estimate the causal effects: the inverse-variance weighted method (IVW); the maximum likelihood estimator computed without knowledge of the correlation between the traits, that is, by setting the off-diagonal entries of the $\Sigma_{Xj}^{*}$ matrices to zero (MLE); and the maximum likelihood estimator computed by incorporating the sample correlation between the traits into these matrices (MLE (cor)). The median estimates across all replications are shown in Figure \ref{fg:bias}. Tables \ref{tb:sims_S1}--\ref{tb:sims_S3} show the coverage and the proportion of replications where the causal null was rejected (which represents the empirical power when the true causal effect was non-zero, and the type I error rate when the true causal effect was null).

In Scenario 1, when there is zero measurement error, IVW provides an unbiased estimate of both causal effects. As the amount of measurement error on $X_{1}$ increases, the median IVW estimate of $\theta_{1}$ moves away from the truth, toward the null. At the same time, the median IVW estimate of $\theta_{2}$ moves away from its true null value. The biases are fairly notable even at relatively small levels of measurement error. As shown in Table \ref{tb:sims_S1}, the coverage of the estimate of $\theta_{1}$ falls below the nominal level of $0.95$, and the type I error rate for the estimate of $\theta_{2}$ is above the significance level of $0.05$, when $\sigma_{\zeta 1}^{2} \geqslant 1.5$.

In Scenario 2, the estimates of both $\theta_{1}$ and $\theta_{2}$ remain unbiased at all levels of measurement error. This agrees with the result given in (\ref{eq:ivw1_bias}) and (\ref{eq:ivw2_bias}), and is because the causal exposure ($X_{2}$) is measured without error, whereas the exposure measured with error ($X_{1}$) is not causal. As shown in Table \ref{tb:sims_S2}, coverage and type I error rates are at their nominal levels in all cases.

In Scenario 3, where both exposures are causal and $X_{2}$ is also measured with error, the estimate of $\theta_{2}$ is biased toward the null and that of $\theta_{1}$ is biased away from the null when $X_{1}$ is measured without error. As the level of measurement error in $X_{1}$ increases, this trend is reversed, so that when $\sigma_{\zeta_{1}}^{2} > \sigma_{\zeta_{2}}^{2}$, $\theta_{1}$ is biased away from the null and $\theta_{2}$ is biased toward the null.

Both MLE methods account for the bias due to measurement error at low levels of correlation between genetic associations ($\rho = 0, 0.2$). At higher levels of correlation ($\rho = 0.4, 0.6$), the MLE estimate is biased, however it retains coverage and type I error rates at their nominal levels. The power of the MLE method is lower, reflecting that it has wider confidence intervals due to accounting for the extra uncertainty in the genetic variant-exposure estimates. When the trait correlation estimates were incorporated, the MLE (cor) remained unbiased, even at high levels of trait correlation. It also retained coverage and type I error rates at their nominal levels, and did not suffer from lower power when compared with IVW.

\begin{table}
	\centering
	\caption{Results from simulations from Scenario 1 where $X_{1}$ is causal ($\theta_{1} = 0.2$) and $X_{2}$ is not causal ($\theta_{2} = 0$) for $Y$, for varying values for the causal effect of $X_{1}$ on $X_{2}$ ($\rho$) and the variance of the measurement error on $X_{1}$ ($\sigma_{\zeta_{1}}^{2}$). For the IVW and MLE methods, and for each of $\theta_{1}$ and $\theta_{2}$, reported is the standard deviation of estimates (SD), coverage (nominal level 0.95), power ($\theta_{1}$) / type I error ($\theta_{2}$) rate at 0.05 significance level (Rej).}
	\label{tb:sims_S1}
	\begin{tabular}{r r c c c c | c c c c | c c c c } \cline{3-14}
		& & \multicolumn{4}{c |}{MR-IVW} & \multicolumn{4}{c |}{MLE} & \multicolumn{4}{c}{MLE (cor)} \\ \cline{3-14}
		& & \multicolumn{2}{c}{$\theta_{1}$} & \multicolumn{2}{c |}{$\theta_{2}$} & \multicolumn{2}{c}{$\theta_{1}$} & \multicolumn{2}{c |}{$\theta_{2}$} & \multicolumn{2}{c}{$\theta_{1}$} & \multicolumn{2}{c}{$\theta_{2}$} \\ \hline
		$\rho$ & $\sigma_{\zeta_{1}}^{2}$ & Cov & Rej & Cov & Rej  & Cov & Rej & Cov & Rej &  Cov & Rej & Cov & Rej \\ \hline
		0&0&0.952&0.895&0.946&0.054&0.939&0.869&0.951&0.049&0.971&0.877&0.950&0.050
		\\
		&1&0.946&0.875&0.946&0.054&0.957&0.847&0.954&0.046&0.976&0.858&0.956&0.044
		\\
		&2&0.927&0.855&0.927&0.073&0.960&0.825&0.959&0.041&0.981&0.839&0.958&0.042
		\\
		&4&0.840&0.793&0.872&0.128&0.956&0.756&0.960&0.040&0.969&0.770&0.957&0.043
		\\ \hline
		0.2&0&0.958&0.736&0.955&0.045&0.953&0.691&0.952&0.048&0.980&0.723&0.952&0.048
		\\
		&1&0.934&0.704&0.932&0.068&0.949&0.645&0.954&0.046&0.967&0.690&0.952&0.048
		\\
		&2&0.927&0.699&0.923&0.077&0.962&0.653&0.955&0.045&0.976&0.680&0.951&0.049
		\\
		&4&0.851&0.635&0.858&0.142&0.946&0.584&0.951&0.049&0.956&0.613&0.946&0.054
		\\ \hline
		0.4&0&0.950&0.606&0.949&0.051&0.951&0.553&0.955&0.045&0.970&0.610&0.947&0.053
		\\
		&1&0.935&0.564&0.931&0.069&0.951&0.504&0.947&0.053&0.975&0.568&0.946&0.054
		\\
		&2&0.913&0.525&0.917&0.083&0.959&0.457&0.967&0.033&0.975&0.507&0.958&0.042
		\\
		&4&0.827&0.434&0.848&0.152&0.971&0.388&0.967&0.033&0.964&0.428&0.955&0.045
		\\ \hline
		0.6&0&0.948&0.418&0.945&0.055&0.948&0.336&0.947&0.053&0.970&0.432&0.947&0.053
		\\
		&1&0.931&0.345&0.932&0.068&0.972&0.279&0.972&0.028&0.958&0.349&0.963&0.037
		\\
		&2&0.892&0.317&0.895&0.105&0.962&0.253&0.963&0.037&0.946&0.320&0.957&0.043
		\\
		&4&0.788&0.235&0.793&0.207&0.962&0.167&0.965&0.035&0.943&0.223&0.956&0.044 \\ \hline
	\end{tabular}
\end{table}

\begin{table}
	\centering
	\caption{Results from simulations from Scenario 1 where $X_{2}$ is causal ($\theta_{2} = 0.2$) and $X_{1}$ is not causal ($\theta_{1} = 0$) for $Y$, for varying values for the causal effect of $X_{1}$ on $X_{2}$ ($\rho$) and the variance of the measurement error on $X_{1}$ ($\sigma_{\zeta_{1}}^{2}$). For the IVW and MLE methods, and for each of $\theta_{1}$ and $\theta_{2}$, reported is the standard deviation of estimates (SD), coverage (nominal level 0.95), power ($\theta_{1}$) / type I error ($\theta_{2}$) rate at 0.05 significance level (Rej).}
	\label{tb:sims_S2}
	\begin{tabular}{r r c c c c | c c c c | c c c c } \cline{3-14}
		& & \multicolumn{4}{c |}{MR-IVW} & \multicolumn{4}{c |}{MLE} & \multicolumn{4}{c}{MLE (cor)} \\ \cline{3-14}
		& & \multicolumn{2}{c}{$\theta_{1}$} & \multicolumn{2}{c |}{$\theta_{2}$} & \multicolumn{2}{c}{$\theta_{1}$} & \multicolumn{2}{c |}{$\theta_{2}$} & \multicolumn{2}{c}{$\theta_{1}$} & \multicolumn{2}{c}{$\theta_{2}$}\\ \hline
		$\rho$ & $\sigma_{\zeta_{1}}^{2}$ & Cov & Rej & Cov & Rej  & Cov & Rej & Cov & Rej &  Cov & Rej & Cov & Rej \\ \hline
		0&0&0.949&0.051&0.952&0.884&0.947&0.053&0.950&0.860&0.976&0.052&0.947&0.866
		\\
		&1&0.946&0.054&0.943&0.893&0.948&0.052&0.950&0.819&0.980&0.052&0.950&0.830
		\\
		&2&0.960&0.040&0.966&0.916&0.960&0.040&0.960&0.801&0.981&0.044&0.961&0.816
		\\
		&4&0.950&0.050&0.953&0.950&0.954&0.046&0.955&0.806&0.973&0.046&0.959&0.825
		\\ \hline
		0.2&0&0.952&0.048&0.959&0.879&0.958&0.042&0.961&0.827&0.986&0.046&0.961&0.859
		\\
		&1&0.948&0.052&0.943&0.905&0.944&0.056&0.947&0.800&0.974&0.051&0.942&0.831
		\\
		&2&0.952&0.048&0.944&0.896&0.958&0.042&0.965&0.750&0.986&0.044&0.963&0.783
		\\
		&4&0.952&0.048&0.948&0.939&0.947&0.053&0.950&0.720&0.975&0.049&0.951&0.753
		\\ \hline
		0.4&0&0.963&0.037&0.958&0.841&0.956&0.044&0.952&0.743&0.980&0.041&0.957&0.796
		\\
		&1&0.949&0.051&0.946&0.873&0.950&0.050&0.942&0.686&0.978&0.053&0.942&0.762
		\\
		&2&0.961&0.039&0.960&0.898&0.960&0.040&0.960&0.660&0.982&0.037&0.957&0.732
		\\
		&4&0.957&0.043&0.954&0.937&0.963&0.037&0.959&0.595&0.976&0.042&0.957&0.649
		\\ \hline
		0.6&0&0.950&0.050&0.945&0.765&0.946&0.054&0.944&0.592&0.969&0.053&0.945&0.720
		\\
		&1&0.962&0.038&0.969&0.831&0.956&0.044&0.955&0.547&0.953&0.037&0.968&0.646
		\\
		&2&0.954&0.046&0.957&0.859&0.957&0.043&0.959&0.474&0.945&0.044&0.954&0.568
		\\
		&4&0.959&0.041&0.955&0.917&0.966&0.034&0.968&0.465&0.947&0.039&0.961&0.542 \\ \hline
	\end{tabular}
\end{table}

\begin{table}
	\centering
	\caption{Results from simulations from Scenario 3 with both $X_{1}$ and $X_{2}$ causal ($\theta_{1} = \theta_{2} = 0.2$) for $Y$ and $X_{2}$ is measured with error, for varying values for the causal effect of $X_{1}$ on $X_{2}$ ($\rho$) and the variance of the measurement error on $X_{1}$ ($\sigma_{\zeta_{1}}^{2}$). For the IVW and MLE methods, and for each of $\theta_{1}$ and $\theta_{2}$, reported is the standard deviation of estimates (SD), coverage (nominal level 0.95), power ($\theta_{1}$) / type I error ($\theta_{2}$) rate at 0.05 significance level (Rej).}
	\label{tb:sims_S3}
	\scalebox{1}{
		\begin{tabular}{r r c c c c | c c c c | c c c c } \cline{3-14}
			& & \multicolumn{4}{c |}{MR-IVW} & \multicolumn{4}{c |}{MLE} & \multicolumn{4}{c}{MLE (cor)} \\ \cline{3-14}
			& & \multicolumn{2}{c}{$\theta_{1}$} & \multicolumn{2}{c |}{$\theta_{2}$} & \multicolumn{2}{c}{$\theta_{1}$} & \multicolumn{2}{c |}{$\theta_{2}$} & \multicolumn{2}{c}{$\theta_{1}$} & \multicolumn{2}{c}{$\theta_{2}$} \\ \hline
			$\rho$ & $\sigma_{\zeta_{1}}^{2}$ & Cov & Rej & Cov & Rej  & Cov & Rej & Cov & Rej &  Cov & Rej & Cov & Rej \\ \hline
			0&0&0.946&0.926&0.946&0.876&0.953&0.768&0.950&0.795&0.980&0.802&0.951&0.819
			\\
			&1&0.957&0.900&0.960&0.909&0.955&0.752&0.959&0.764&0.981&0.776&0.959&0.793
			\\
			&2&0.951&0.879&0.957&0.945&0.962&0.700&0.965&0.744&0.978&0.725&0.965&0.772
			\\
			&4&0.907&0.855&0.918&0.960&0.956&0.657&0.957&0.675&0.977&0.698&0.958&0.707
			\\ \hline
			0.2&0&0.945&0.776&0.944&0.852&0.940&0.549&0.940&0.746&0.967&0.631&0.940&0.791
			\\
			&1&0.949&0.770&0.944&0.903&0.947&0.553&0.946&0.664&0.971&0.631&0.946&0.731
			\\
			&2&0.943&0.771&0.954&0.938&0.955&0.517&0.957&0.641&0.977&0.591&0.956&0.706
			\\
			&4&0.896&0.665&0.908&0.971&0.956&0.456&0.957&0.614&0.970&0.522&0.959&0.675
			\\ \hline
			0.4&0&0.953&0.688&0.950&0.818&0.953&0.409&0.951&0.589&0.976&0.537&0.954&0.725
			\\
			&1&0.954&0.624&0.953&0.892&0.961&0.344&0.957&0.558&0.976&0.452&0.957&0.673
			\\
			&2&0.948&0.578&0.953&0.933&0.960&0.308&0.963&0.489&0.972&0.407&0.964&0.603
			\\
			&4&0.887&0.523&0.894&0.973&0.955&0.294&0.958&0.461&0.964&0.378&0.956&0.536
			\\ \hline
			0.6&0&0.960&0.508&0.944&0.705&0.954&0.223&0.946&0.416&0.973&0.374&0.943&0.599
			\\
			&1&0.955&0.445&0.952&0.828&0.957&0.179&0.961&0.359&0.954&0.319&0.958&0.509
			\\
			&2&0.936&0.376&0.935&0.923&0.958&0.161&0.961&0.355&0.937&0.271&0.959&0.497
			\\
			&4&0.879&0.349&0.878&0.981&0.971&0.130&0.970&0.337&0.942&0.220&0.974&0.447 \\ \hline
		\end{tabular}
	}
\end{table}

Tables S1 and S2 in the Supporting Information give the mean F statistics across each replication for the corresponding scenarios shown in Tables \ref{tb:sims_S1}--\ref{tb:sims_S3}. These values give an indication of the extent of the measurement error on each exposure. F statistics are commonly used to measure the strength of the association between genetic variants and the exposures, and are used because there is a direct relationship between the F statistics and bias due to weak instruments \citep{Zhao2020, Sanderson2021}. We report, for each exposure, two F statistics. The first is the unconditional F statistic obtained from the regression of the exposure on the genetic variants. The second is the conditional F statistic, which assesses the strength of the instruments on one exposure after accounting for the effects of the genetic variants on the other exposure. In order to compute the conditional F statistic for $X_{1}$, the residuals are taken from the regression of $X_{1}$ on the fitted values from the regression of $X_{2}$ on the genetic variants. The conditional F statistic is then the F statistic from the regression of these residuals on the genetic variants. As shown in the tables, the F statistics related to $X_{1}$ decrease as the amount of measurement error increases. In Scenarios 1 and 2, where $X_{2}$ is measured without error, the F statistics related to $X_{2}$ remain at fairly constant levels.

In a supplementary study, we repeated the simulation described above but where $Y$ was a binary outcome. All simulation parameters were the same as before, but $Y$ was generated from the Bernoulli distribution with probability $\theta_{0} + \theta_{1} X_{1} + \theta_{2} X_{2} + U$, where $\theta_{0} = -4$ in Scenarios 1 and 2, and $\theta_{0} = -4.9$ in Scenario 3. These values for $\theta_{0}$ were chosen so that the outcome occurred with approximately $0.05$ prevalence. The results of these simulations are shown in Figure S1 and Tables S3--S5 in the Supporting Information. The effects of measurement error on the causal effect estimates, and the ability of the MLE methods to correct for it, are broadly in line with the primary simulation study.

\subsection{Estimated proportion of effect mediated}
In order to demonstrate the effect of measurement error on estimating the proportion of the total effect of an exposure mediated by another variable, we simulated from the same model as in the previous section with $\theta_{2} = 0.2$ and $\rho = 0.6$. The exposure of interest $X_{1}$ had a causal effect of either $\theta_{1} =0$ or $0.1$ and was measured without error, with $\sigma_{\zeta_{1}}^{2} = 0$, or with error, with $\sigma_{\zeta_{1}}^{2} = 1$. The amount of measurement error on $X_{2}$ was varied by setting $\sigma_{\zeta_{2}}^{2}$ between $0$ and $4$. The genetic associations for the first $10$ variants with $X_{2}$ were set to zero, so that these variants were valid instrumental variables for estimating the total effect of $X_{1}$ on $Y$ in a univariable analysis.

\begin{figure}
	\centering
	\includegraphics{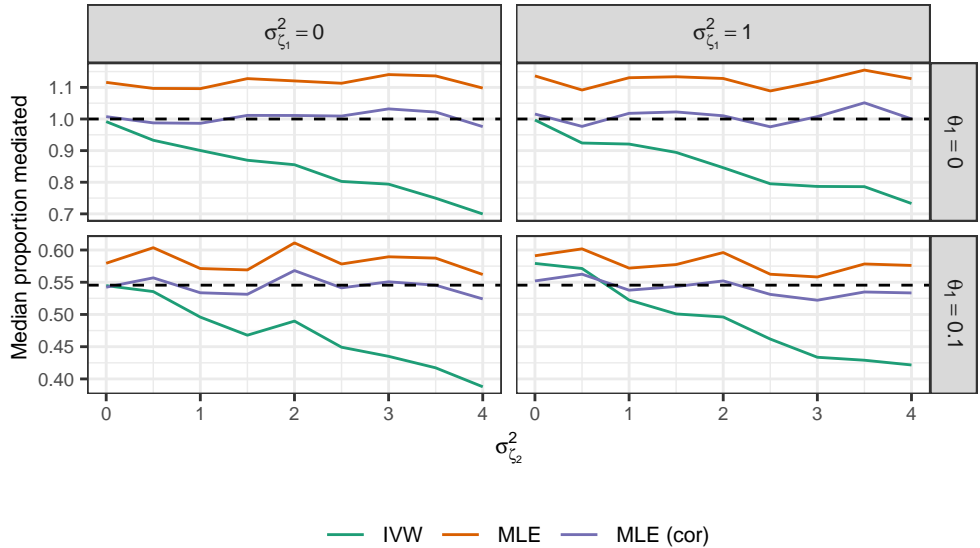}
	\caption{Median estimated proportion of the effect of $X_{1}$ on $Y$ mediated by $X_{2}$ for varying values of the direct effect ($\theta_{1}$), the variance of the measurement error on $X_{1}$ ($\sigma_{\zeta_{1}}^{2}$) and the variance of the measurement error on $X_{2}$ ($\sigma_{\zeta_{2}}^{2}$), using IVW and maximum likelihood estimation without (MLE) and with (MLE (cor)) sample trait correlation included. The dashed black lines indicate the true proportion mediated.}
	\label{fg:pmed}
\end{figure}

The proportion mediated was estimated by $1$ minus the ratio of the causal effect estimate of $X_{1}$ on $Y$ in a multivariable analysis (using either IVW, MLE or MLE (cor)) to the causal effect estimate in a univariable analysis. For the univariable analyses, only the first $10$ genetic variants were used as instruments. Figure \ref{fg:pmed} plots the median estimated proportion mediated for each scenario and value of $\sigma_{\zeta_{2}}^{2}$.

In all cases, the estimated proportion using IVW became more underestimated as the amount of measurement error on $X_{2}$ increased. The MLE method showed a small level of bias when trait correlation was not taken into account, but this bias did not increase as $\sigma_{\zeta_{2}}^{2}$ increased. When trait correlation estimates were included, the MLE method produced unbiased estimates of the true proportion mediated at all levels of measurement error.

\section{Applied examples}

\subsection{The effect of body mass index and waist-to-hip ratio on heart disease}
We considered the effect of body mass index (BMI) and waist-to-hip ratio (WHR) on risk of coronary heart disease (CHD). BMI and WHR are both known risk factors for heart disease \citep{VanGaal_2006}. Anthropometric traits are known to be prone to measurement error, with waist and hip circumference measures more prone than height and weight measures \citep{Ulijaszek1999}. We applied univariable analyses, which considered the effects of BMI and WHR separately on CHD, and multivariable analyses which included both exposures in the same model. For the univariable analyses, we took as instruments genetic variants which were associated with the exposure at genome-wide significance (p-value $< 5 \times 10^{-8}$) from the GWAS of \citet{Pulit2018}. These variants were pruned to have $r^2 < 0.001$. For the multivariable analyses, we took as instruments genetic variants which were associated with either BMI or WHR and pruned to have $r^2 < 0.001$. In the multivariable analysis, the combined list of genetic variants were pruned with respect to their association with BMI. The associations of the genetic instruments with CHD were taken from the GWAS of \cite{Nikpay2015} and accessed using PhenoScanner \citep{Phenoscanner1, Phenoscanner2}.

\begin{figure}
	\centering
	\includegraphics{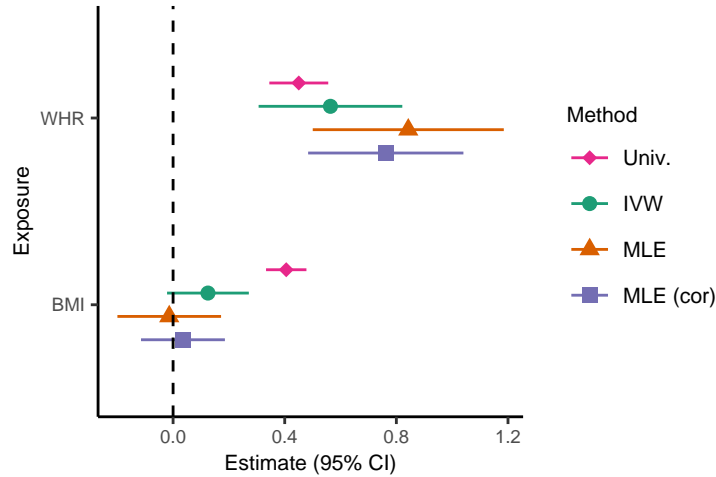}
	\caption{Log odds ratio for CHD and 95\% confidence intervals per standard deviation increase in BMI and WHR using: univariable Mendelian randomization (Univ.); the inverse-variance weighted method (IVW); and the maximum likelihood method without (MLE) and with (MLE (cor)) sample trait correlation incorporated.}
	\label{fg:bmiwhr}
\end{figure}

We estimated the effects of the exposures on CHD using IVW (univariable and multivariable cases) and the MLE method (multivariable case only). We applied the MLE method both without trait correlation incorporated, as well as with the full $\Sigma_{Xj}^{*}$ matrices estimated using the correlation between BMI and WHR in the UK Biobank dataset, which is $0.433$ \citep{Pulit2018}. The results are show in Figure \ref{fg:bmiwhr}.

In the univariable analyses, both exposures showed a positive effect on CHD risk, with 95\% confidence intervals above the null in both cases. The log odds ratio of CHD per standard deviation increase in BMI was 0.405 (95\% confidence interval 0.333, 0.478), and per standard deviation increase in WHR was 0.450 (0.345, 0.556). In the multivariable IVW analyses, the effect of WHR remained strongly positive, with log odds ratio of 0.564 (0.306, 0.821), but the effect of BMI attenuated toward the null, with log odds ratio of 0.125 (-0.022, 0.271). One possible explanation for this is that some of the effect of BMI on CHD is mediated by WHR. It may also be the case that the genetic variants which associate with BMI, but not WHR, have pleiotropic effects. When the MLE method was used, the effect of BMI attenuated further to the null, with log odds ratio (when incorporating sample trait correlation) of 0.035 (-0.115, 0.186), whereas the effect of WHR increased, with log odds ratio of 0.762 (0.484, 1.040). This is consistent with a scenario where WHR is measured with error and is causal, whereas BMI is measured without error and is not causal conditional on WHR. This is plausible given that WHR is more prone to measurement error than BMI. If this were the case, because the exposures are correlated, the IVW analysis would bias BMI away from the null and WHR toward the null, as in Scenario 1 of the simulation study.

\subsection{The portion of the effect of education on heart disease mediated by other factors} \label{se:edu}
\cite{Carter2019} considered the effect of educational attainment on cardiovascular disease, and the portion of this effect which was mediated by BMI, a lifetime smoking score and systolic blood pressure (SBP). Using both multivariable regression with observational data and Mendelian randomization, they showed that a large proportion of the effect of educational attainment on CHD is mediated by these three other exposures (estimates of 42\% in regression analysis and  36\% in Mendelian Randomization analysis). We considered this example to assess whether measurement error on the mediators may lead to underestimation of this proportion.
	
We performed multivariable Mendelian randomization with genetic association estimates taken from the GWAS of \citet{Lee2018} (educational attainment), \cite{Pulit2018} (BMI), \citet{Wootton2020} (lifetime smoking score), the Neale Lab (http://www.nealelab.is/uk-biobank/) (SBP) and \citet{Nikpay2015} (CHD). The genetic association estimates with educational attainment and CHD were taken from the supplementary tables supplied by \citet{Carter2019}, and with the other traits from their respective GWAS results. We used as instruments the genetic variants which were identified by \citet{Carter2019} as being associated with at least one of the exposures at genome-wide significance and pruned to have $r^{2} < 0.001$. We performed univariable Mendelian randomization using as instruments the genetic variants identified by \citet{Carter2019} as being associated with educational attainment at genome-wide significance and pruned to have $r^{2} < 0.001$. Note that since we are using some genetic association estimates from different GWAS results and a different pruning threshold for the univariable analysis, we do not expect to replicate the results of \citet{Carter2019}. Rather, we aim to demonstrate how the estimated proportion of the mediated effect of educational attainment on CHD may be affected by measurement error.

Table \ref{tb:edu} shows the estimated total (from univariable analyses) and direct (from multivariable analyses) effects of educational attainment on CHD using both the IVW and MLE methods. Also shown are the corresponding estimates of the proportion of the effect which is mediated by the other exposures. Confidence intervals for these estimates were constructed using standard errors computed using the delta method (see Section S3 in the Supporting Information). Using IVW, which ignores measurement error, it is estimated that 54.8\% (95\% confidence interval 25.1, 84.5\%) of the effect of educational attainment on CHD is mediated. However, this estimate rises to 66.4\% (35.1, 97.7\%) when measurement error is incorporated using the MLE approach.

\begin{table}
	\centering
	\caption{Estimated total effect and direct effect (expressed as log odds ratios, with 95\% confidence intervals) of educational attainment on coronary heart disease, as well as the estimated proportion of the total effect mediated by BMI, smoking and SBP, using both the IVW and MLE methods.}
	\label{tb:edu}
	\begin{tabular}{l c c} \hline
		& IVW & MLE\\ \hline
		Total effect & -0.481 (-0.584, -0.378) & -0.493 (-0.583, -0.402) \\
		Direct effect & -0.218 (-0.353, -0.083) & -0.165 (-0.317, -0.014) \\ \hline
		Proportion mediated & 0.548 (0.251, 0.845) & 0.664 (0.351, 0.977) \\ \hline
	\end{tabular}
\end{table}

\section{Discussion}
In this paper, we have shown that multivariable Mendelian randomization studies which use genetic associations with exposures measured with error are subject to bias, and that this bias may be in any direction. Furthermore, exposures which are measured with error may bias the effect estimates of other exposures with which they are correlated, even if these other exposures are measured precisely. As a result, effect estimates from multivariable Mendelian randomization analyses cannot be assumed to be conservative, that is, biased toward the null, as they can in the single exposure case. Furthermore, we have shown that in mediation analyses, biased effect estimates due to measurement error may lead to substantially incorrect estimates of the relative contribution to a causal effect from potential mediators.

The simulation study demonstrated that the bias from measurement error can start to impact the conclusions from Mendelian randomization studies at relatively modest levels. Furthermore, this cannot always be diagnosed using conventional F statistic thresholds. For example, in simulation Scenarios 1 and 2, the F statistics for the exposures (both conditional and unconditional) were the same. This is because the only differences in these scenarios was the causal effects of the exposures on the outcome. Nonetheless, while in Scenario 1, both effect estimates from IVW showed considerable bias, loss of power ($\theta_{1}$) and inflated type I errors ($\theta_{2}$), the estimates in Scenario 2 were unbiased and retained correct coverage and type I errors. Thus, the F statistics do not necessarily indicate whether bias from measurement error is a problem or not.

We have shown that measurement error on exposures can be accounted for by incorporating the genetic variant-exposure covariance matrices in the estimator using a maximum likelihood framework, and that this approach retains correct coverage and type I error rates. This is true even if the exposures are correlated but this correlation is not incorporated into the estimates. That is, if only the diagonal elements of the covariance matrices, taken from GWAS summary results, for example, are input. If estimates are available of the correlation between the exposures, the maximum likelihood estimator can incorporate this information, and there is only minimal loss of power compared with IVW. In implementing this method, we do not need to model the measurement error or estimate its magnitude.

The bias due to classical measurement error can be thought of in a similar way to weak instrument bias in Mendelian randomization studies, since the consequence of this type of measurement error is to increase the standard errors of the genetic association estimates. Methods for accounting for weak instrument bias in multivariable Mendelian randomization have previously been considered by \citet{Sanderson2021} and \cite{Wang2021}. Their proposed estimators are similar to the MLE proposed here, although they also include a heterogeneity parameter to account for balanced pleiotropy. Our approach assumes that all instruments are valid and so does not attempt to model pleiotropy. As long as this assumption is true, the MLE estimator will be more efficient than those which include a heterogeneity parameter. Furthermore, the iterative algorithm we propose for computing the MLE is computationally fast and easy to implement, and the asymptotic variance allows for confidence intervals to be computed. One potential extension to the MLE approach is to include a random effects adjustment, which is typically used in IVW to account for heterogeneity in the individual variant ratio estimates \citep{Thompson1999}.

Although summary level Mendelian randomization can be performed using genetic associations with both continuous or binary outcomes, it should be noted that our theoretical results will not apply in the binary case. This is due to the non-collapsibility of odds ratios \citep{Vansteelandt2011}. Nonetheless, our supplementary simulation showed that the effects of measurement error on IVW estimation still broadly follow the same pattern, and that the MLE approach is still able to provide less biased estimates with nominal coverage and type I error rates.

Not considered here is systematic measurement error which leads to biased genetic association estimates. Some of this type of error is typically accounted for in GWAS by including technical covariates such as assessment centre in the genetic variant-trait regressions. Nonetheless, for certain traits, some systematic measurement error may remain. Bias in genetic association estimates will lead to bias in Mendelian randomization estimates. If some prior knowledge could be assumed as to the nature of this systematic bias, one approach to account for it could be via a Bayesian framework, where a prior distribution is placed on the mean genetic association estimate. This is an area for future research.

Overall, we have demonstrated that measurement error on exposures can affect multivariable Mendelian randomization estimates in realistic settings and that, unlike in the single exposure case, the resulting bias will not necessarily tend toward the null. We have also examined the implications of measurement error when using multivariable Mendelian randomization for mediation analysis. We have proposed an algorithm for causal effect estimation derived in a maximum likelihood framework which can account for this type of measurement error. This provides an important sensitivity analysis in a multivariable Mendelian randomization study.

\section*{Software}
IVW analyses were performed in R using the MendelianRandomization packge, version 0.5.0 \citep{Yavorska2017,Broadbent2020}. Pruning of genetic variants was performed in R using the TwoSampleMR package, version 0.5.5 \citep{twosamplemr}.

\section*{Acknowledgments}
AJG and SB are supported by a Sir Henry Dale Fellowship jointly funded by the Wellcome Trust and the Royal Society (grant number 204623/Z/16/Z). This research was funded by the NIHR Cambridge Biomedical Research Centre (BRC-1215-20014). The views expressed are those of the authors and not necessarily those of the NHS, the NIHR or the Department of Health and Social Care.

\bibliographystyle{apalike_ag.bst}
\bibliography{MVMR_ME_ref}


\end{document}


\title{Supporting information for: Bias in multivariable Mendelian randomization studies due to measurement error on exposures}
\author[1]{Jiazheng Zhu}
\author[1,2]{Stephen Burgess}
\author[1]{Andrew J. Grant\thanks{Corresponding author. Email address: andrew.grant@mrc-bsu.cam.ac.uk}}
\affil[1]{\normalsize MRC Biostatistics Unit, University of Cambridge, Cambridge, UK}
\affil[2]{\normalsize Cardiovascular Epidemiology Unit, University of Cambridge, Cambridge, UK}
\date{}

\maketitle

\setcounter{table}{0}
\setcounter{figure}{0}
\setcounter{equation}{0}
\renewcommand{\thesection}{S}
\renewcommand{\thetable}{S\arabic{table}}
\renewcommand{\thefigure}{S\arabic{figure}}
\renewcommand{\theequation}{S\arabic{equation}}
\renewcommand{\thesubsection}{S\arabic{subsection}}

\subsection{Measurement error in the two exposure case}
From (4) and (5), we have
\begin{align*}
\hat{\theta}_{IVW}^{*} - \theta = - \left( \sum_{j=1}^{J} \sigma_{Yj}^{-2} \hat{\beta}^{*}_{Xj} \hat{\beta}^{*'}_{Xj} \right)^{-1} \sum_{j=1}^{J} \sigma_{Yj}^{-2} \left( \varepsilon_{Yj} \beta_{Xj} + \varepsilon_{Yj} \hat{\beta}_{\zeta_{j}} - \beta_{Xj} \hat{\beta}_{\zeta_{j}}' \theta - \hat{\beta}_{\zeta_{j}} \hat{\beta}_{\zeta_{j}}' \theta \right) .
\end{align*}
Now, the random variables $\sigma_{Y1}^{-1} \varepsilon_{Y1}, \ldots, \sigma_{YJ}^{-1} \varepsilon_{YJ}$ are independent and identically distributed (i.i.d.) with mean zero. Furthermore, the random variables $\sigma_{Y1}^{-1} \hat{\beta}_{\zeta_{1k}}, \ldots, \sigma_{YJ}^{-1} \hat{\beta}_{\zeta_{Jk}}$ are i.i.d. with mean zero, since $\sigma_{Yj}^{-1} \textrm{Var} \left( \hat{\beta}_{\zeta_{jk}} \right)$ is constant for all $j, k$. Finally, note that $ \sigma_{Yj}^{-1} \varepsilon_{Yj}$, $\sigma_{Yj}^{-1} \hat{\beta}_{\zeta_{jk}}$ are independent for all $j, k$. Thus, denoting the $k$th element of the vector $a$ by $\left\lbrace a \right\rbrace _{k}$,
\begin{align*}
\left\lbrace J^{-1} \sum_{j=1}^{J} \sigma_{Yj}^{-2} \left( \varepsilon_{Yj} \beta_{Xj} \right) \right\rbrace_{k} &= J^{-1} \sum_{j=1}^{J} \left( \sigma_{Yj}^{-1} \varepsilon_{Yj} \right) \left( \sigma_{Yj}^{-1} \beta_{Xjk} \right) \rightarrow 0 \\
\left\lbrace J^{-1} \sum_{j=1}^{J} \sigma_{Yj}^{-2} \left( \varepsilon_{Yj} \hat{\beta}_{\zeta j} \right) \right\rbrace_{k} &= J^{-1} \sum_{j=1}^{J} \left( \sigma_{Yj}^{-1} \varepsilon_{Yj} \right) \left( \sigma_{Yj}^{-1} \hat{\beta}_{\zeta jk} \right) \rightarrow 0 \\
\left\lbrace J^{-1} \sum_{j=1}^{J} \sigma_{Yj}^{-2} \left( \beta_{Xj} \hat{\beta}_{\zeta j}' \theta \right) \right\rbrace_{k} &= J^{-1} \sum_{j=1}^{J} \sum_{l=1}^{K} \left( \sigma_{Yj}^{-1} \beta_{X jk} \right) \left( \sigma_{Yj}^{-1} \hat{\beta}_{\zeta jl} \right) \theta_{l} \rightarrow 0 ,\\
\end{align*}
as $J \rightarrow \infty$. Therefore, from (6) and (7),
\[
\hat{\theta}_{IVW}^{*} - \theta \rightarrow \begin{pmatrix} v_{X1}^{*} & c_{X}^{*} \\ c_{X}^{*} & v_{X2}^{*} \end{pmatrix}^{-1} \begin{pmatrix} v_{\zeta 1} \theta_{1} \\ v_{\zeta 2} \theta_{2} \end{pmatrix} ,
\]
and the result follows.

\subsection{Asymptotic variance of the MLE}
Denote by $\hat{\theta}$ the maximiser of $\tilde{l} \left( \theta \right)$ and by $\theta_{0}$ the true value of $\theta$. Assuming that $\hat{\theta}$ is a consistent estimator of $\theta_{0}$ then, using the standard Taylor series expansion argument, $\hat{\theta} - \theta_{0}$ is asymptotically normally distributed with mean zero and variance $\mathcal{I}^{-1} \Omega \mathcal{I}^{-1}$, where
\[
\mathcal{I} = E \left\lbrace - \frac{\partial^2 \tilde{l}\left( \theta_{0} \right)}{\partial \theta \partial \theta'} \right\rbrace, \quad \Omega = \textrm{var} \left\lbrace \frac{\partial \tilde{l}\left( \theta_{0} \right)}{\partial \theta} \right\rbrace .
\]
Let
\begin{align*}
v_{j} &= \sigma_{Yj}^{2} + \theta' \Sigma_{Xj}^{*} \theta \\
e_{j} &= \hat{\beta}_{Yj} - \hat{\beta}_{Xj}^{* \prime} \theta \\
g_{j} \left( \theta \right) &= v_{j}^{-2} e_{j}.
\end{align*}
The $k$th element of $\partial g_{j} \left( \theta \right) / \partial \theta$ is
\[
v_{j}^{-1/2} \hat{\beta}_{Xjk}^{*} + v_{j}^{-3/2} e_{j} \left\lbrace \Sigma_{Xj}^{*} \theta \right\rbrace_{k} ,
\]
where $\left\lbrace a \right\rbrace_{k}$ denotes the $k$th element of vector $a$. The ($k$, $l$)th element of  $\partial^2 g_{j} \left( \theta \right) / \partial \theta \partial \theta'$ is
\[
v_{j}^{-3/2} \left[ -2 \left\lbrace \Sigma_{Xj}^{*} \theta \right\rbrace_{l} \hat{\beta}_{Xjk}^{*} + \left\lbrace \Sigma_{Xj}^{*} \theta \right\rbrace_{k} \hat{\beta}_{Xjl}^{*} - e_{j} \Sigma_{Xjkl}^{*} \right] + 3 v_{j}^{-5/2} \left[ v_{j} \hat{\beta}_{Xjk}^{*} + e_{j} \left\lbrace \Sigma_{Xj}^{*} \theta \right\rbrace_{k} \right] \left\lbrace \Sigma_{Xj}^{*} \theta \right\rbrace_{l} .
\]
The optimisation given in (11) is equivalent to the estimator proposed by \cite{Wang2021} with no heterogeneity parameter and a quadratic loss function. Following \cite{Wang2021}, we can therefore estimate $\Omega$ by
\[
\sum_{j=1}^{p} \frac{\partial g_{j} \left( \hat{\theta} \right)}{\partial \theta} \frac{\partial g_{j} \left( \hat{\theta} \right)}{\partial \theta'}
\]
and we can estimate $\mathcal{I}$ by
\[
\sum_{j=1}^{p} \left\lbrace \frac{\partial g_{j} \left( \hat{\theta}\right)}{\partial \theta} \frac{\partial g_{j} \left( \hat{\theta} \right)}{\partial \theta'} + g_{j} \left( \hat{\theta} \right) \frac{\partial^2 g_{j} \left( \hat{\theta} \right)}{\partial \theta \partial \theta'} \right\rbrace .
\]

\subsection{Standard error of the estimated proportion mediated}
Denote the causal effect estimate of $X_{1}$ on $Y$ from a univariable analysis by $\hat{\theta}_{UV}$ with standard error $\sigma_{UV}$, and from a multivariable analysis by $\hat{\theta}_{MV}$ with standard error $\sigma_{MV}$. The true values that these analyses estimate are $\theta_{UV}$ and $\theta_{MV}$, respectively. We assume that $\left( \hat{\theta}_{MV}, \hat{\theta}_{UV} \right)'$ is normally distributed with mean $\left( \theta_{MV}, \theta_{UV} \right)'$ and covariance matrix
\[
\begin{pmatrix}
\sigma_{MV}^{2} & \sigma_{MV}^{2} \sigma_{MV}^{2} \tilde{\rho} \\
\sigma_{MV}^{2} \sigma_{MV}^{2} \tilde{\rho} & \sigma_{UV}^{2} 
\end{pmatrix},
\]
where $\tilde{\rho}$ is the correlation between $\hat{\theta}_{MV}$ and $\hat{\theta}_{UV}$. By the delta method, $\hat{\theta}_{MV} / \hat{\theta}_{MV}$ is approximately normally distributed with mean $\theta_{MV} / \theta_{MV}$ and variance
\begin{equation}
\frac{\sigma_{MV}^{2}}{\theta_{UV}^{2}} + \frac{\theta_{MV}^{2} \sigma_{UV}^{2}} {\theta_{UV}^{4}} - 2 \frac{\theta_{MV} \sigma_{MV}^{2} \sigma_{MV}^{2} \tilde{\rho}}{\theta_{UV}^{3}}. \label{eq:delta}
\end{equation}
In the applied example presented in Section 5.2, we estimate the standard error of the proportion mediated by plugging in the estimates of the parameters in the first two terms in (\ref{eq:delta}). This assumes that the estimators are uncorrelated, which will only be true in the null case of no mediation. Thus, our standard error is overestimated, and hence the confidence intervals given may be too wide. Nonetheless, this is a conservative approach which should ensure coverage is over the nominal level.

\bibliographystyle{apalike_ag.bst}
\bibliography{MVMR_ME_ref}

\begin{table}[p]
	\centering
	\caption{F statistics (F) and conditional F statistics (CF) for Scenarios 1 and 2 where $X_{1}$ is measured with error and $X_{2}$ is measured without error, for varying values for the causal effect of $X_{1}$ on $X_{2}$ ($\rho$) and the variance of the measurement error on $X_{1}$ ($\sigma_{\zeta_{1}}^{2}$).}
	\begin{tabular}{r r c c | c c } \cline{3-6}
		& & \multicolumn{2}{c |}{F} & \multicolumn{2}{c}{Conditional F} \\ \cline{1-6}
		$\rho$ & $\sigma_{\zeta_{1}}^{2}$ & $X_{1}$ & $X_{2}$ & $X_{1}$ & $X_{2}$ \\ \hline
		0&0&142.45&142.01&17.08&17.04
\\
		&1&72.05&142.83&9.05&17.95
\\
		&2&48.34&143.09&6.40&18.95
\\
		&4&29.27&142.20&4.24&20.62
\\ \hline
		0.2&0&141.97&191.84&12.11&16.41
\\
		&1&71.85&191.87&6.53&17.45
\\
		&2&48.41&192.48&4.66&18.55
\\
		&4&29.12&192.14&3.19&21.11
\\ \hline
		0.4&0&142.34&233.22&8.58&14.08
\\
		&1&71.91&233.52&4.77&15.53
\\
		&2&48.26&234.14&3.49&16.98
\\
		&4&29.23&233.17&2.49&19.89
\\ \hline
		0.6&0&142.33&261.07&5.78&10.62
\\
		&1&71.93&260.51&3.38&12.26
\\
		&2&48.37&260.47&2.57&13.86
\\
		&4&29.35&261.19&1.94&17.29\\ \hline
	\end{tabular}
\end{table}

\begin{table}[p]
	\centering
	\caption{F statistics (F) and conditional F statistics (CF) for Scenario 3 where $X_{1}$ and $X_{2}$ are both measured without error, for varying values for the causal effect of $X_{1}$ on $X_{2}$ ($\rho$) and the variance of the measurement error on $X_{1}$ ($\sigma_{\zeta_{1}}^{2}$).}
	\begin{tabular}{r r c c | c c } \cline{3-6}
		& & \multicolumn{2}{c |}{F} & \multicolumn{2}{c}{Conditional F} \\ \cline{1-6}
		$\rho$ & $\sigma_{\zeta_{1}}^{2}$ & $X_{1}$ & $X_{2}$ & $X_{1}$ & $X_{2}$ \\ \hline
		0&0&142.54&71.82&18.09&9.10
\\
		&1&71.81&71.52&9.50&9.45
\\
		&2&48.11&71.68&6.66&9.95
\\
		&4&29.31&71.76&4.36&10.67
\\ \hline
		0.2&0&142.98&97.48&12.60&8.60
\\
		&1&71.61&96.85&6.87&9.30
\\
		&2&48.41&96.98&4.91&9.85
\\
		&4&29.32&97.26&3.32&11.02
\\ \hline
		0.4&0&142.58&118.96&9.14&7.64
\\
		&1&71.43&118.27&4.99&8.28
\\
		&2&48.28&118.52&3.65&8.98
\\
		&4&29.22&118.91&2.59&10.58
\\ \hline
		0.6&0&142.90&133.18&6.33&5.91
\\
		&1&71.78&132.91&3.61&6.71
\\
		&2&48.04&133.00&2.75&7.63
\\
		&4&29.13&132.86&2.03&9.31 \\ \hline
	\end{tabular}
\end{table}

\begin{figure}[p]
	\centering
	\includegraphics{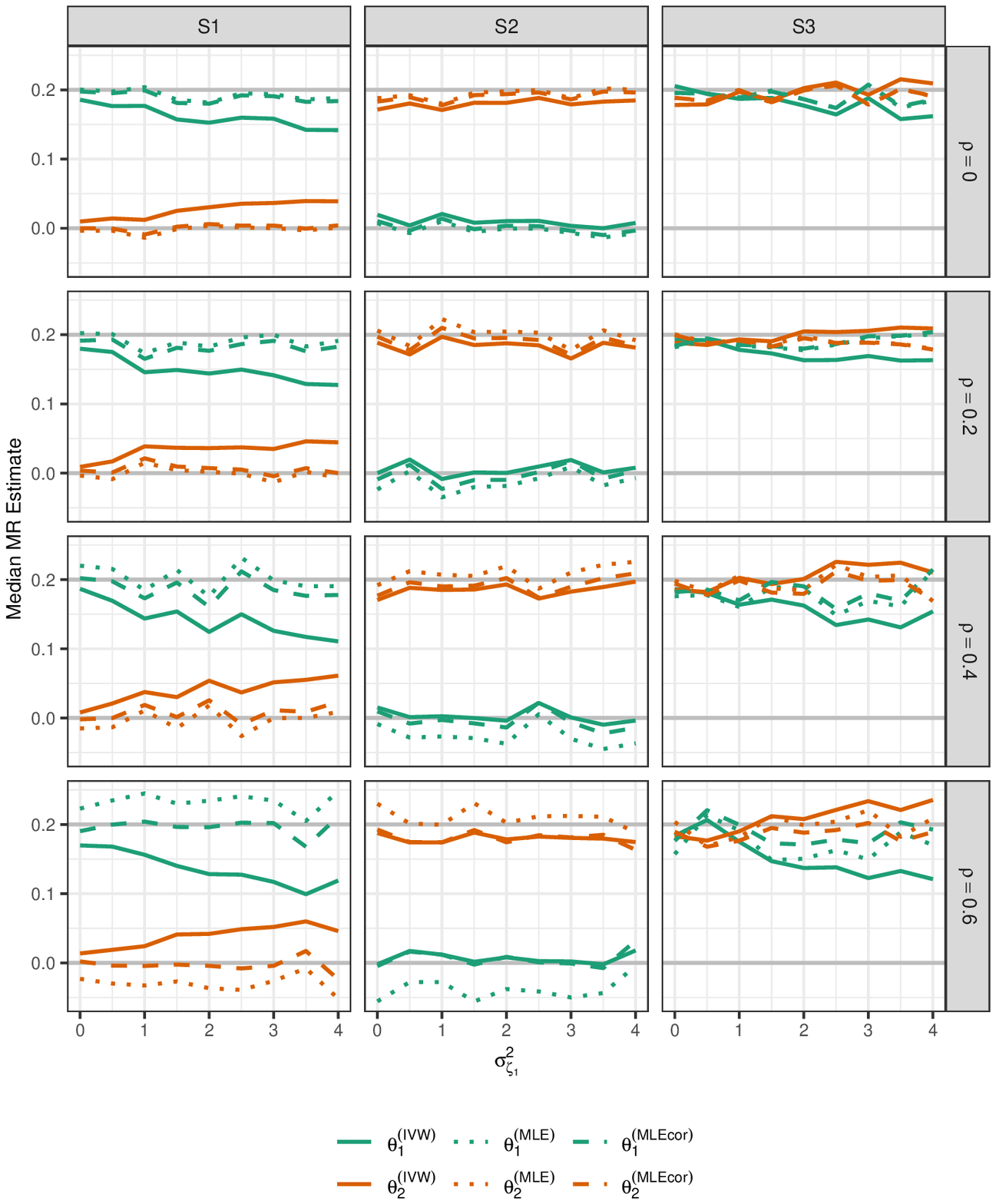}
	\caption{Median estimates of the causal effects from simulation Scenarios 1--3 with a binary outcome, for varying values for the causal effect of $X_{1}$ on $X_{2}$ ($\rho$) and the variance of the measurement error on $X_{1}$ ($\sigma_{\zeta_{1}}^{2}$), using IVW ($\theta_{1}^{\textrm{IVW}}$, $\theta_{2}^{\textrm{IVW}}$) and maximum likelihood estimation without ($\theta_{1}^{\textrm{MLE}}$, $\theta_{2}^{\textrm{MLE}}$) and with ($\theta_{1}^{\textrm{MLEcor}}$, $\theta_{2}^{\textrm{MLEcor}}$) sample trait correlation included. The thick grey lines indicate $0.2$ and $0$, which are the true causal effects in the various settings.}
\end{figure}

\begin{table}[p]
	\centering
	\caption{Results from simulations from Scenario 1 where $X_{1}$ is causal ($\theta_{1} = 0.2$) and $X_{2}$ is not causal ($\theta_{2} = 0$) for binary outcome $Y$, for varying values for the causal effect of $X_{1}$ on $X_{2}$ ($\rho$) and the variance of the measurement error on $X_{1}$ ($\sigma_{\zeta_{1}}^{2}$). For the IVW and MLE methods, and for each of $\theta_{1}$ and $\theta_{2}$, reported is the standard deviation of estimates (SD), coverage (nominal level 0.95), power ($\theta_{1}$) / type I error ($\theta_{2}$) rate at 0.05 significance level (Rej).}
	\begin{tabular}{r r c c c c | c c c c | c c c c } \cline{3-14}
		& & \multicolumn{4}{c |}{MR-IVW} & \multicolumn{4}{c |}{MLE} & \multicolumn{4}{c}{MLE (cor)} \\ \cline{3-14}
		& & \multicolumn{2}{c}{$\theta_{1}$} & \multicolumn{2}{c |}{$\theta_{2}$} & \multicolumn{2}{c}{$\theta_{1}$} & \multicolumn{2}{c |}{$\theta_{2}$} & \multicolumn{2}{c}{$\theta_{1}$} & \multicolumn{2}{c}{$\theta_{2}$} \\ \hline
		$\rho$ & $\sigma_{\zeta_{1}}^{2}$ & Cov & Rej & Cov & Rej  & Cov & Rej & Cov & Rej &  Cov & Rej & Cov & Rej \\ \hline
		0&0&0.952&0.162&0.958&0.042&0.948&0.158&0.953&0.047&0.971&0.163&0.951&0.049
\\
		0&1&0.955&0.150&0.965&0.035&0.954&0.152&0.962&0.038&0.976&0.164&0.962&0.038
\\
		0&2&0.945&0.141&0.956&0.044&0.950&0.140&0.956&0.044&0.978&0.144&0.954&0.046
\\
		0&4&0.951&0.127&0.944&0.056&0.956&0.132&0.951&0.049&0.971&0.135&0.951&0.049
\\ \hline
		0.2&0&0.950&0.111&0.949&0.051&0.945&0.105&0.949&0.051&0.965&0.114&0.946&0.054
\\
		0.2&1&0.956&0.106&0.959&0.041&0.953&0.109&0.955&0.045&0.969&0.119&0.958&0.042
\\
		0.2&2&0.949&0.117&0.952&0.048&0.953&0.114&0.948&0.052&0.971&0.121&0.946&0.054
\\
		0.2&4&0.943&0.095&0.950&0.050&0.969&0.085&0.965&0.035&0.979&0.092&0.965&0.035
\\ \hline
		0.4&0&0.956&0.098&0.951&0.049&0.958&0.093&0.951&0.049&0.972&0.102&0.947&0.053
\\
		0.4&1&0.963&0.070&0.963&0.037&0.967&0.067&0.969&0.031&0.978&0.075&0.964&0.036
\\
		0.4&2&0.951&0.077&0.952&0.048&0.960&0.071&0.959&0.041&0.973&0.077&0.957&0.043
\\
		0.4&4&0.929&0.068&0.935&0.065&0.959&0.061&0.957&0.043&0.967&0.068&0.951&0.049
\\ \hline
		0.6&0&0.950&0.070&0.953&0.047&0.951&0.065&0.951&0.049&0.962&0.072&0.946&0.054
\\
		0.6&1&0.964&0.071&0.967&0.033&0.969&0.060&0.967&0.033&0.962&0.075&0.965&0.035
\\
		0.6&2&0.939&0.070&0.942&0.058&0.949&0.062&0.957&0.043&0.949&0.075&0.950&0.050
\\
		0.6&4&0.944&0.065&0.940&0.060&0.968&0.054&0.964&0.036&0.942&0.063&0.960&0.040 \\ \hline
	\end{tabular}
\end{table}

\begin{table}[p]
	\centering
	\caption{Results from simulations from Scenario 2 where $X_{2}$ is causal ($\theta_{2} = 0.2$) and $X_{1}$ is not causal ($\theta_{1} = 0$) for binary outcome $Y$, for varying values for the causal effect of $X_{1}$ on $X_{2}$ ($\rho$) and the variance of the measurement error on $X_{1}$ ($\sigma_{\zeta_{1}}^{2}$). For the IVW and MLE methods, and for each of $\theta_{1}$ and $\theta_{2}$, reported is the standard deviation of estimates (SD), coverage (nominal level 0.95), power ($\theta_{1}$) / type I error ($\theta_{2}$) rate at 0.05 significance level (Rej).}
	\begin{tabular}{r r c c c c | c c c c | c c c c } \cline{3-14}
		& & \multicolumn{4}{c |}{MR-IVW} & \multicolumn{4}{c |}{MLE} & \multicolumn{4}{c}{MLE (cor)} \\ \cline{3-14}
		& & \multicolumn{2}{c}{$\theta_{1}$} & \multicolumn{2}{c |}{$\theta_{2}$} & \multicolumn{2}{c}{$\theta_{1}$} & \multicolumn{2}{c |}{$\theta_{2}$} & \multicolumn{2}{c}{$\theta_{1}$} & \multicolumn{2}{c}{$\theta_{2}$} \\ \hline
		$\rho$ & $\sigma_{\zeta_{1}}^{2}$ & Cov & Rej & Cov & Rej  & Cov & Rej & Cov & Rej &  Cov & Rej & Cov & Rej \\ \hline
		0&0&0.946&0.054&0.953&0.156&0.945&0.055&0.947&0.156&0.975&0.055&0.946&0.160
\\
		0&1&0.967&0.033&0.955&0.131&0.960&0.040&0.954&0.123&0.988&0.042&0.952&0.129
\\
		0&2&0.954&0.046&0.949&0.173&0.951&0.049&0.951&0.149&0.977&0.051&0.949&0.153
\\
		0&4&0.968&0.032&0.964&0.195&0.968&0.032&0.960&0.139&0.984&0.031&0.961&0.143
\\ \hline
		0.2&0&0.951&0.049&0.945&0.190&0.947&0.053&0.946&0.174&0.969&0.054&0.943&0.192
\\
		0.2&1&0.958&0.042&0.959&0.190&0.955&0.045&0.953&0.157&0.974&0.045&0.953&0.167
\\
		0.2&2&0.956&0.044&0.948&0.180&0.953&0.047&0.952&0.144&0.978&0.047&0.949&0.155
\\
		0.2&4&0.967&0.033&0.964&0.201&0.965&0.035&0.968&0.117&0.982&0.035&0.966&0.123
\\ \hline
		0.4&0&0.951&0.049&0.946&0.160&0.949&0.051&0.948&0.141&0.972&0.058&0.944&0.157
\\
		0.4&1&0.965&0.035&0.963&0.189&0.960&0.040&0.960&0.144&0.974&0.043&0.958&0.158
\\
		0.4&2&0.948&0.052&0.950&0.194&0.951&0.049&0.947&0.138&0.971&0.052&0.945&0.160
\\
		0.4&4&0.954&0.046&0.954&0.229&0.960&0.040&0.962&0.117&0.968&0.043&0.959&0.130
\\ \hline
		0.6&0&0.955&0.045&0.957&0.145&0.952&0.048&0.951&0.122&0.960&0.051&0.948&0.141
\\
		0.6&1&0.950&0.050&0.954&0.140&0.957&0.043&0.956&0.095&0.956&0.051&0.950&0.110
\\
		0.6&2&0.943&0.057&0.952&0.168&0.946&0.054&0.953&0.099&0.952&0.062&0.947&0.121
\\
		0.6&4&0.965&0.035&0.954&0.175&0.964&0.036&0.962&0.083&0.932&0.039&0.955&0.095 \\ \hline
	\end{tabular}
\end{table}

\begin{table}[p]
	\centering
	\caption{Results from simulations from Scenario 3 with both $X_{1}$ and $X_{2}$ causal ($\theta_{1} = \theta_{2} = 0.2$) for binary outcome $Y$ and $X_{2}$ is measured with error, for varying values for the causal effect of $X_{1}$ on $X_{2}$ ($\rho$) and the variance of the measurement error on $X_{1}$ ($\sigma_{\zeta_{1}}^{2}$). For the IVW and MLE methods, and for each of $\theta_{1}$ and $\theta_{2}$, reported is the standard deviation of estimates (SD), coverage (nominal level 0.95), power ($\theta_{1}$) / type I error ($\theta_{2}$) rate at 0.05 significance level (Rej).}
	\begin{tabular}{r r c c c c | c c c c | c c c c } \cline{3-14}
		& & \multicolumn{4}{c |}{MR-IVW} & \multicolumn{4}{c |}{MLE} & \multicolumn{4}{c}{MLE (cor)} \\ \cline{3-14}
		& & \multicolumn{2}{c}{$\theta_{1}$} & \multicolumn{2}{c |}{$\theta_{2}$} & \multicolumn{2}{c}{$\theta_{1}$} & \multicolumn{2}{c |}{$\theta_{2}$} & \multicolumn{2}{c}{$\theta_{1}$} & \multicolumn{2}{c}{$\theta_{2}$} \\ \hline
		$\rho$ & $\sigma_{\zeta_{1}}^{2}$ & Cov & Rej & Cov & Rej  & Cov & Rej & Cov & Rej &  Cov & Rej & Cov & Rej \\ \hline
		0&0&0.951&0.156&0.952&0.140&0.945&0.117&0.951&0.136&0.972&0.123&0.951&0.141
\\
		0&1&0.955&0.159&0.958&0.164&0.955&0.129&0.954&0.132&0.972&0.132&0.952&0.136
\\
		0&2&0.958&0.149&0.958&0.182&0.953&0.115&0.953&0.119&0.975&0.119&0.953&0.124
\\
		0&4&0.951&0.150&0.947&0.223&0.955&0.121&0.948&0.134&0.976&0.124&0.947&0.139
\\ \hline
		0.2&0&0.958&0.136&0.959&0.153&0.954&0.100&0.960&0.131&0.973&0.112&0.960&0.138
\\
		0.2&1&0.956&0.126&0.958&0.172&0.955&0.102&0.956&0.117&0.972&0.110&0.954&0.135
\\
		0.2&2&0.957&0.110&0.957&0.210&0.958&0.086&0.955&0.123&0.978&0.093&0.954&0.139
\\
		0.2&4&0.935&0.115&0.943&0.236&0.954&0.092&0.951&0.111&0.972&0.099&0.949&0.118
\\ \hline
		0.4&0&0.959&0.120&0.957&0.153&0.954&0.081&0.955&0.117&0.965&0.101&0.955&0.137
\\
		0.4&1&0.962&0.095&0.960&0.185&0.957&0.067&0.957&0.119&0.976&0.084&0.954&0.135
\\
		0.4&2&0.951&0.101&0.964&0.216&0.955&0.072&0.962&0.109&0.964&0.091&0.961&0.126
\\
		0.4&4&0.944&0.092&0.965&0.227&0.959&0.064&0.961&0.085&0.963&0.072&0.960&0.098
\\ \hline
		0.6&0&0.945&0.096&0.948&0.109&0.945&0.068&0.946&0.070&0.966&0.083&0.945&0.101
\\
		0.6&1&0.962&0.076&0.962&0.152&0.964&0.050&0.967&0.075&0.949&0.062&0.962&0.104
\\
		0.6&2&0.939&0.103&0.949&0.194&0.947&0.065&0.955&0.084&0.940&0.088&0.948&0.104
\\
		0.6&4&0.946&0.065&0.952&0.274&0.965&0.043&0.965&0.084&0.942&0.052&0.962&0.102 \\ \hline
	\end{tabular}
\end{table}